\documentclass[11pt]{article}
\usepackage{moriond,epsfig}

\def\NCA{{\em Nuovo Cimento}}

\def\PRL{{\em Phys. Rev. Lett.}}
\def\PRD{{\em Phys. Rev.} D}

\def\be{\begin{equation}}
\def\ee{\end{equation}}
\def\bea{\begin{eqnarray}}
\def\eea{\end{eqnarray}}

\begin{document}
\vspace*{4cm}
\title{DO THE $\pi N$ TOTAL CROSS SECTIONS INCREASE LIKE
log $\nu$ OR (log $\nu )^2$\\  AT HIGH ENERGIES ?\ 
\footnote{Talk presented at Rencontres de Moriond on QCD and
Hadronic Interactions, March 16-23, 2002}}

\author{(Presented by Keiji Igi)\\
 K. Igi and M. Ishida$^*$}

\address{Department of Information Science, Kanagawa University, Hiratsuka\\
Kanagawa 259-1293, Japan\\
$^*$Department of Physics, Tokyo Institute of Technology\\
Tokyo 152-8551, Japan}

\maketitle\abstracts{
We propose to use rich informations on $\pi p$ total cross sections below 
$N(\sim 10$ GeV) in order to investigate whether these cross sections increase
like log $\nu$ or (log $\nu )^2$ at high energies. A finite-energy sum rule (FESR)
which is derived in the spirit of the $P^\prime$ sum rule as well as the $n=1$ moment
FESR have been required to constrain the high-energy parameters. We then searched
for the best fit of $\sigma_{\rm tot}^{(+)}$ above 70 GeV 
in terms of high-energy parameters
constrained by these two FESR. We can conclude that our analysis strongly favours
the (log $\nu )^2$ behaviors satisfying the Froissart unitarity bound.}

As is well-known, the sum of $\pi^+ p$ and $\pi^- p$ total cross sections
has a tendency to increase above 70 GeV 
experimentally\cite{rf1}. 
It has not been known\cite{Cudell}, however, if this increase behaves like log $\nu$
or log$^2$ $\nu$ consistent with the Froissart bound.\cite{rf2}

We would like to propose to use rich informations of  $\pi p$
total cross sections at low and intermediate energy regions in order to 
investigate the high energy behaviours of $\pi p$ total cross sections above
70 GeV using new finite-energy sum rules (FESR) as constraints.
   
Such a kind of attempt has been initiated in Ref.~\cite{rf3}.
The $s$-wave $\pi N$ scattering length $a^{(+)}$ of the crossing-even amplitude had been
expressed as 
\begin{eqnarray}
\left(  1+\frac{\mu}{M}  \right) a^{(+)} &=& -\frac{g_r^2}{4\pi}\left( \frac{\mu}{2M} \right)^2
\frac{1}{M} \frac{1}{1-(\frac{\mu}{2M})^2} +\frac{1}{2\pi^2}\int_0^\infty dk 
[\sigma_{\rm tot}^{(+)}(k)-\sigma_{\rm tot}^{(+)}(\infty )]\ \ \ \ \ \ \ \ \ \ 
\label{eq1}
\end{eqnarray}
with pion mass $\mu$ under the assumption that there are no singularities 
with the vacuum quantum numbers in the $J$ plane except for the Pomeron $(P)$.
The evidence that this sum rule had not been satisfied led us to the prediction of the 
$P^\prime$ trajectory with $\alpha_{P^\prime}\approx 0.5$, and 
soon the $f$ meson ($f_2(1275)$)
has been uncovered on this $P^\prime$ trajectory.\\

\hspace*{-0.8cm}(\underline{\it FESR(1)}): Taking into account 
the present situation of increasing total cross section data,
we derive FESR in the spirit of the $P^\prime$ sum rule\cite{rf3}. 
We consider the 
crossing-even (spin-averaged) forward scattering amplitude for $\pi p$ scattering\cite{rf4} 
\begin{eqnarray}
f^{(+)}(\nu ) &=& \frac{1}{4\pi} [ A^{(+)}(\nu )+\nu B^{(+)}(\nu ) ] .
\label{eq2}
\end{eqnarray}
We assume
\begin{eqnarray}
{\rm Im}\  f^{(+)}(\nu )  & \simeq & {\rm Im}\ R(\nu )+{\rm Im}\ f_{P^\prime}(\nu ) \nonumber\\
 &=& \frac{\nu}{\mu^2}\left( c_0+c_1 {\rm log}\ \frac{\nu}{\mu} 
+ c_2 {\rm log}^2\ \frac{\nu}{\mu}    \right) 
+\frac{\beta_{P^\prime}}{\mu} \left( \frac{\nu}{\mu} \right)^{\alpha_{P^\prime}}
\label{eq3}
\end{eqnarray}
at high energies $(\nu \geq N)$. 
Taking into account the amplitude $f^{(+)}(\nu )$ to be crossing-even.
we can derive (for a detail see ref. \cite{rfIM} )
\begin{eqnarray}
{\rm Re}\ f^{(+)}(\mu ) &=& {\rm Re}\ R(\mu) + {\rm Re}\ f_{P^\prime}(\mu) 
-\frac{g_r^2}{4\pi}\left(\frac{\mu}{2M}\right)^2\frac{1}{M}\frac{1}{1-(\frac{\mu}{2M})^2} \nonumber\\
&&+\frac{1}{2\pi^2}\int_0^{\overline{N}} \sigma_{\rm tot}^{(+)}(k) dk
-\frac{2P}{\pi}\int_0^N \frac{\nu}{k^2}
\left\{ 
{\rm Im}\ R(\nu )+\frac{\beta_{P^\prime}}{\mu}
\left(\frac{\nu}{\mu}\right)^{0.5}  \right\} d\nu \ ,\ \ \ \ \ \ \ \ \ \ \ 
\label{eq8} 
\end{eqnarray}
where  $\overline{N}\equiv \sqrt{N^2-\mu^2} \simeq N$. 
Let us call Eq.~(\ref{eq8}) as the FESR(1) which we use as the first 
constraint. It is important to notice that Eq.~(\ref{eq8}) reduces to 
the $P^\prime$ sum rule in ref.\cite{rf3} if $c_1,\ c_2\rightarrow 0$.

The FESR (\cite{rf5}, \cite{rf6}, \cite{rf7})
\begin{eqnarray}
\int_0^N d\nu \ \nu^n {\rm Im}\ f(\nu ) &=& \sum_i \beta_i \frac{N^{\alpha_i+n+1}}{\alpha_i+n+1}
\label{eq88}
\end{eqnarray}
holds for even positive integer $n$ when $f(\nu )$ is crossing odd,
and holds for odd positive integer $n$ when $f(\nu )$ is crossing even.
We can also derive negative-integer moment FESR.
The only significant FESR is a one for $f^{(+)}(\nu )/\nu $ corresponding to $n=-1$.
FESR(1) belongs to this case.\\

\hspace*{-0.8cm}(\underline{\it FESR(2)}): The second FESR corresponding to $n=1$ is:
\begin{eqnarray} 
\pi\mu \left( \frac{g_r^2}{4\pi} \right) \left( \frac{\mu}{2M} \right)^3
 &+&  \frac{1}{4\pi}\int_0^{\overline{N}}dk\ k^2\sigma_{\rm tot}^{(+)}(k) \nonumber\\
 &=& \int_0^N \nu {\rm Im}\ R(\nu ) d\nu + \int_0^N \nu {\rm Im}\ f_{P^\prime}(\nu )d\nu\ \ . 
\label{eq9}
\end{eqnarray}
We call Eq.~(\ref{eq9}) as the FESR(2). It is to be noticed that the contribution from
higher energy regions is enhanced. \\

\hspace*{-0.8cm}(\underline{\it Data})\ \ The numerical value of 
$-\frac{g_r^2}{4\pi}\left(\frac{\mu}{2M}\right)^2
\frac{1}{M}\frac{1}{1-(\frac{\mu}{2M})^2}
=-0.0854$GeV$^{-1}$, $\pi \mu \frac{g_r^2}{4\pi} \left( \frac{\mu}{2M} \right)^3=0.0026$GeV
have been evaluated using $\frac{g_r^2}{4\pi}=14.4$.
Re $f^{(+)}(\mu )=\left( 1+\frac{\mu}{M} \right) a^{(+)}
=\left( 1+\frac{\mu}{M} \right) \frac{1}{3}(a_{\frac{1}{2}}+2a_{\frac{3}{2}})
=-(0.014\pm 0.026)$GeV$^{-1}$ was obtained from\cite{rf8} 
$a_{\frac{1}{2}}=(0.171\pm 0.005)\mu^{-1}$
and $a_{\frac{3}{2}}=-(0.088\pm 0.004)\mu^{-1}$.

We have used rich data\cite{rf1} of $\sigma^{\pi^+ p}$ and $\sigma^{\pi^- p}$
to evaluate the relevant integrals of cross sections
appearing in FESR(1) and (2). We have obtained 
${\displaystyle \frac{1}{2\pi^2}\int_0^{\overline{N}} dk\ 
\sigma_{\rm tot}^{(+)}(k) }=38.75\pm0.25$ GeV$^{-1}$, 
${\displaystyle \frac{1}{4\pi}\int_0^{\overline{N}} dk\ k^2 
\sigma_{\rm tot}^{(+)}(k) }=1817\pm$31 GeV for $\overline{N}=10$ GeV. 
For a detail, see ref. \cite{rfIM}.\\

\hspace*{-0.8cm}(\underline{\it Analysis})\ \  The FESR(1) and (2) are our 
starting points. Armed with these two, we expressed high-energy parameters 
$c_0$, $c_1$, $c_2$, 
$\beta_{P^\prime}$ in terms of the Born term and the $\pi N$ scattering length $a^{(+)}$
as well as the total cross sections up to $N$. We then attempt to 
fit the $\sigma_{\rm tot}^{(+)}$ above 70GeV. We set $N=10GeV$.

Let us first define the log$^2\nu$ model and the log $\nu$ model.
The log$^2\nu$ model is a model for which the imaginary part of $f^{(+)}(\nu )$
behaves as $a+b\ {\rm log}\ \nu+c({\rm log}\ \nu)^2$ as $\nu$ becomes large\cite{KN75}.
The log $\nu$ model is a model for which the imaginary part of $f^{(+)}(\nu )$
behaves as $a^\prime +b^\prime\ {\rm log}\ \nu$ for large $\nu$.
So we generally assume that the Im $f^{(+)}(\nu )$ behaves as Eq.~(\ref{eq3})
at high energies $(\nu\geq N)$.\\
{\bf (1) log $\nu$ model}:\ \ This model has three parameters 
$c_0$, $c_1$ and $\beta_{P^\prime}$ with two constraints FESR (1), (2).
We set $N=10$GeV and expressed
both $c_0$, $\beta_{P^\prime}$ as a function of $c_1$ using the FESR(1) and (2).
We obtained 
\begin{eqnarray}
c_0(c_1) &=& 0.0879-4.94c_1,\ \ \beta_{P^\prime}(c_1) = 0.1290-7.06c_1\ .
\label{eq12}
\end{eqnarray}  
We then tried to fit 12 data points of $\sigma_{\rm tot}^{(+)}(k)$ 
between 70GeV and 340GeV. The best fit we obtained is $c_1=0.00185$ 
which gives $c_0=0.0787$ and $\beta_{P^\prime}=0.142$ with the bad 
``reduced $\chi^2$," 
$\chi^2/(N_{\rm data}-N_{\rm param})=29.03/(12-1)\simeq 2.6$.
Therefore it turned out that this model has difficulties to reproduce 
the experimental increase of $\pi p$ total cross sections above 70GeV.

\hspace*{-0.8cm}{\bf (2) log$^2$ $\nu$ model}:\ \  
This model has four parameters $c_0$, $c_1$, $c_2$ and $\beta_{P^\prime}$ 
with two constraints FESR(1),(2). 
We again set $N=10$GeV and required both FESR(1) and (2) as constraints.
Then $c_0$, $\beta_{P^\prime}$ are expressed as functions of $c_1$ and $c_2$ as
\begin{eqnarray}
c_0(c_1,c_2) &=& 0.0879-4.94c_1-21.50c_2,\ \ 
\beta_{P^\prime}(c_1,c_2) = 0.1290-7.06c_1-41.46c_2 .\ \ \ \ \ \ \ \ \ \ \ 
\label{eq13}
\end{eqnarray}  
We then searched for the fit to 12 data points of $\sigma_{\rm tot}^{(+)}(k)$
above 70GeV. The best fit in terms of two parameters $c_1$ and $c_2$ led us 
to greatly improved value of ``reduced $\chi^2$," 
$\chi^2/(N_{\rm data}-N_{\rm param})=0.746/(12-2)\simeq 0.075$ for
$c_1=-0.0215<0$ and $c_2=0.00182>0$ which give 
$c_0=0.155$ and $\beta_{P^\prime}=0.0574$. 
This is an excellent fit to the data.

It is remarkable to notice that the wide range of data ($k\geq 5$GeV) 
have been reproduced within the error even in the region where the fit 
has not been made (see Fig. 1 (a) and (b)). It is also important to 
note that the results do not change so much for the value of $N$. 
The increase of $\sigma_{\rm tot}^{(+)}$ above 50 GeV
is explained via log$^2$ $\nu/\mu$ $(c_2>0)$ and the decrease between 
$5\sim 50$GeV is explained by log $\nu/\mu$ ($c_1<0$).

Therefore, we can conclude that our analysis based on the FESR(1),(2) 
strongly favours
the  log$^2$ $\nu/\mu$ behaviours satisfying the 
Froissart unitarity bound.\\

\hspace*{-0.8cm}\underline{\it Note added in proof.}-- After completing 
the manuscript, we were informed by Dr. Jurgen Engelfried that the SELEX 
collaboration, U.~Dersch et al. $[$ Nucl.~Phys.~{\bf B579} (2000) 277 $]$ 
had a datum for $\pi^- N$ at 610 GeV. Our log$^2\ \nu$ model predicts 
25.9mb for $\sigma_{\rm tot}^{(+)}$ at 610GeV which is consistent with 
their value on $\pi^- N$, $(26.6\pm 0.9)$mb .
We were also informed by Dr. Bararab Nicolescu that COMPETE collaboration,
J.~R.~Cudell et al. $[$hep-ph/0107219$]$ also reached a similar conclusion that the
Froissart bound seemed favoured by completely different approach. We also
came to know from a talk at the 37th Moriond Conference (March 16-23, 2002)
by Dr.~F.~D.~Steffen that the gluon saturation leads to log$^2\ \nu$ behaviours 
at high energy $[$ A.~I.~Shoshi, F.~D.~Steffen and H.~J.~Pirner, 
hep-ph/0202012 $]$.

\begin{figure}
  \epsfxsize=14. cm
  \centerline{\epsffile{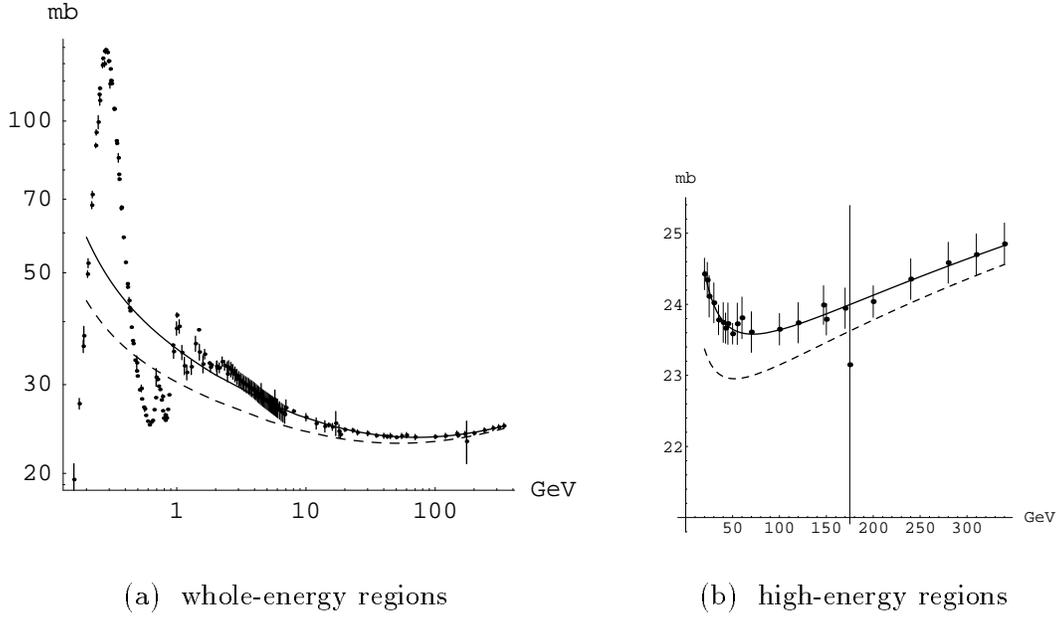}}
  \caption{Fit to the $\sigma_{\rm tot}^{(+)}$ data above 70GeV 
by the log$^2$ $\nu$ model.
The dashed line represents the contribution from Im $R(\nu )$ with $c_2>0$.}
\end{figure}


\newpage

\section*{References}
\begin{thebibliography}{99}
\bibitem{rf1} Particle Data Group, D. E. Groom et al., 
{\em Eur. Phys. J. C}  {\bf 15} (2000) 235.
\bibitem{Cudell} J.~R.~Cudell et al., \PRD\ {\bf 61} (2000) 034019. 
\bibitem{rf2} M. Froissart, {\em Phys. Rev.}\ {\bf 123} (1961) 1053.\\
A.~Martin, \NCA\ {\bf 42} (1966) 930.
\bibitem{rf3} K.~Igi, \PRL\ {\bf 9} (1962) 76.
\bibitem{rf4} G.~F.~Chew, M.~L.~Goldberger, F.~E.~Low and Y.~Nambu,
{\em Phys.~Rev.}~{\bf 106} (1957) 1337.
\bibitem{rfIM} K.~Igi and M.~Ishida, hep-ph/0202163.
\bibitem{rf5} K.~Igi and S.~Matsuda, \PRL\ {\bf 18} (1967) 625.
\bibitem{rf6} A.~A.~Logunov, L.~D.~Soloviev and A.~N.~Tavkhelidze, 
{\em Phys.~Lett.}~{\bf 24B} (1967) 181.
\bibitem{rf7} R.~Dolen, D.~Horn and C.~Schmid, 
\PRL\ {\bf 19} (1967) 402; {\em Phys.~Rev.}~{\bf 166} (1968) 1768.
\bibitem{rf8} R.~K.~Bhaduri,``Models of the Nucleon," 
Addison Wesley pub., 1988, p.134.
\bibitem{KN75} K.~Kang and B.~Nicolescu, \PRD\ {\bf 11} (1975) 2461.
\end{thebibliography}
\end{document}